\documentclass[manuscript]{acmart}

\usepackage{dirtytalk}

\usepackage{graphicx}
\usepackage{subfigure}
\usepackage{multirow}
\usepackage{calc}

\usepackage{textcomp}
\usepackage{siunitx}
\usepackage{soul,xcolor}
\usepackage{gensymb}

\AtBeginDocument{%
  \providecommand\BibTeX{{%
    \normalfont B\kern-0.5em{\scshape i\kern-0.25em b}\kern-0.8em\TeX}}}

\newcommand{\purple}[1]{\textcolor{purple}{#1}}

\begin{document}

\setstcolor{red}



\title[Is it me?]{Is it Me? Toward Self-Extension to AI Avatars in Virtual Reality}

\author{Jieying Zhang}
\orcid{0000-0002-4881-1350}
\affiliation{%
 \institution{University of Amsterdam}
 \city{Amsterdam}
 \country{The Netherlands}}
\affiliation{%
 \institution{Centrum Wiskunde \& Informatica}
 \city{Amsterdam}
 \country{The Netherlands}}
\email{jieying.zhang@student.uva.nl}

\author{Steeven Villa}
\orcid{0000-0002-4881-1350}
\affiliation{%
  \institution{LMU Munich}
  \city{Munich}
  \country{Germany}
}
\email{villa@posthci.com}

\author{Abdallah El Ali}
\orcid{0000-0002-9954-4088}
\affiliation{%
  \institution{Centrum Wiskunde \& Informatica}
    \city{Amsterdam}
    \country{The Netherlands}}
  \affiliation{%
  \institution{Utrecht University}
   \city{Utrecht}
  \country{The Netherlands}
}
\email{aea@cwi.nl}

 \renewcommand{\shortauthors}{Cond. accepted to CHI '26 EA}

\begin{abstract}

Advances in generative AI, speech synthesis, and embodied avatars enable systems that not only assist communication, but can act as proxies on users’ behalf. Prior work in HCI has largely focused on systems as external tools, with less attention paid to the experiential consequences of users’ speech and actions becoming assimilated with AI-generated output. We introduce the design and implementation of ProxyMe, a work-in-progress VR prototype that allows users to embody an avatar whose voice and spoken content are modified by an AI system. By combining avatar-based embodiment, voice cloning, and AI-mediated speech augmentation, ProxyMe invites the exploration of avatar self-extension: situations in which AI-modified communication is experienced as part of one’s own expressive behavior. We chart out research challenges and envisioned scenarios, with a focus on how varying degrees of delegation and steerability can influence perceived agency, authorship, and self-identification. 



\end{abstract}

\begin{teaserfigure}
       \centering
       \setlength{\abovecaptionskip}{2pt}
       \includegraphics[width=0.9\textwidth]{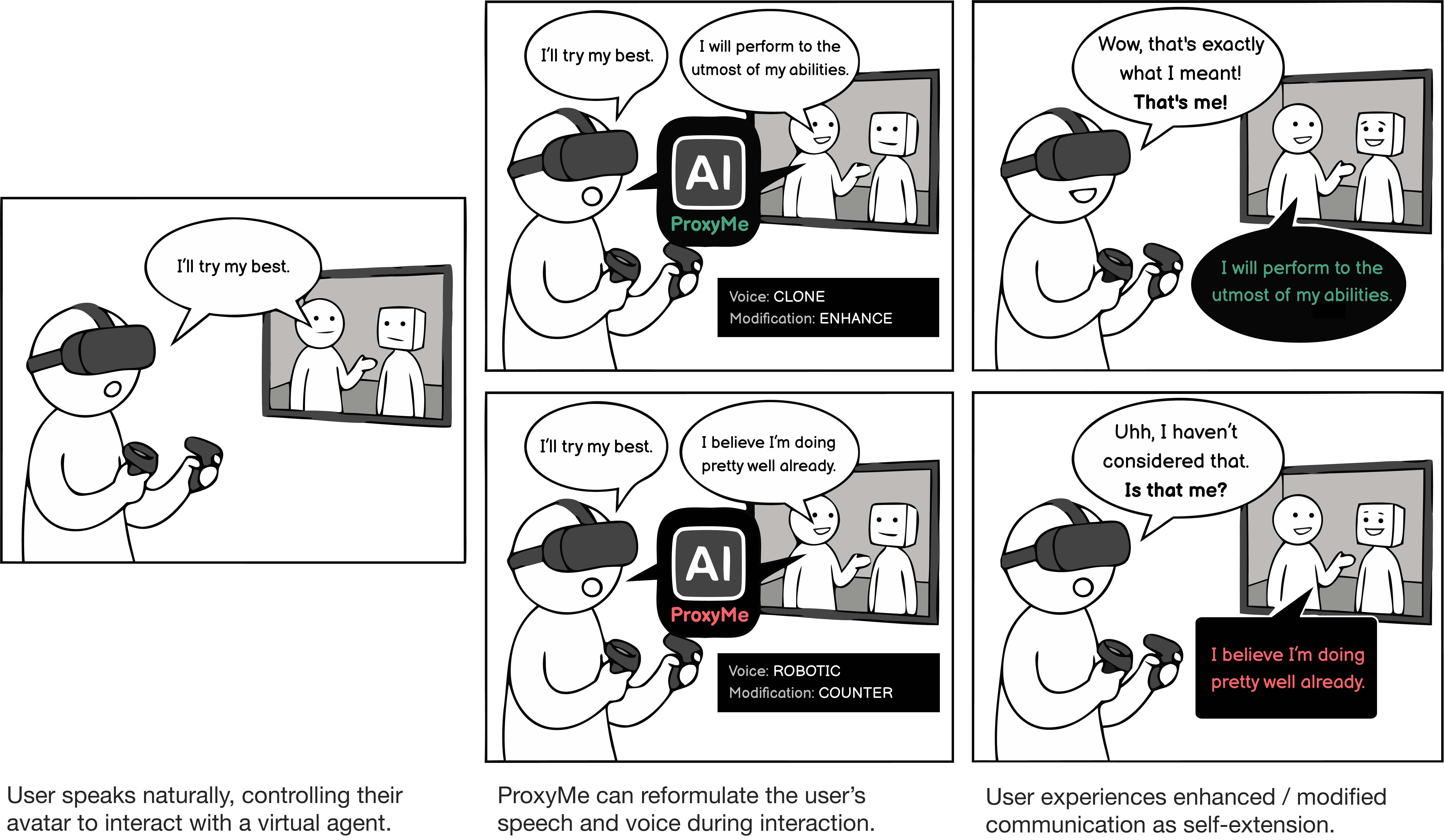}
               \caption{ProxyMe interaction concept.}                
       \label{fig:teaser}
       \Description{A user embodies in a VR avatar, interacts with a virtual agent, and gives an initial response (e.g., I'll try my best). ProxyMe takes the avatar's utterance as input and generates a verbal continuation after the virtual agent asks for more details. Experimental conditions manipulate continuation along two factors: (1) voice (cloned vs. robotic) and (2) content modification (repetition, enhancement, or counter-conclusion), enabling tests of how vocal and content variation shape perceived agency, authorship, and embodiment in AI-mediated co-speech.} 
   \end{teaserfigure}

\ccsdesc[500]{Human-centered computing~Virtual reality}
\ccsdesc[500]{Human-centered computing~Human computer interaction (HCI)}

\keywords{Self-extension, avatar, human augmentation, AI-mediated communication, voice cloning}



\maketitle






\section{Introduction}


Science fiction narratives have envisioned the possibility of humans that are no longer confined to their physical body, allowing them the ability to extend their cognitive and bodily self onto external substrates. Ghost in the Shell \cite{wikighostintheshell} depicts \textit{cyberbrains} and \textit{prosthetic bodies (shells)} as mechanisms through which cognition and action are distributed beyond the biological organism, which raises questions about the continuity of identity and the attribution of actions to the self.
By contrast, Ergo Proxy \cite{wikiergoproxy} portrays \textit{“proxies”} as substitutive agents that act in place of human subjects, often preceding conscious intention or endorsement. Rather than extending human agency transparently, these agents displace authorship and decision-making, which ultimately expose the fragility of human agency and responsibility under such mediated contexts.

Recent advances in generative models, 3D avatars, and speech synthesis bring these aforementioned science fiction visions into experimentally tractable forms. This is especially due to such capabilities embedded into real-time and multi-modal communication contexts such as voice calls, video meetings, and immersive VR environments \cite{yangStreamVCRealTimeLowLatency2024, zhangCospeechVideoGeneration2025}. Furthermore, contemporary AI agents are capable of generating persuasive arguments \cite{salviConversationalPersuasivenessGPT42025}, supporting reasoning and decision-making on users' behalf \cite{chernyakovAIEthicsWhen2022, danryWearableReasonerEnhanced2020,chengConversationalAgentsYour2025}, or allowing fine-grained vocal control or even full voice cloning \cite{fangLeveragingAIGeneratedEmotional2025}. Such developments give rise to applications in which users may co-speak with, or be represented by, AI systems, blurring boundaries between self and simulation, and as far as authenticity and algorithmic generation \cite{kourosDigitalMirrorsAI2024, zhengLearningAIClonesEnhancing2025}. Such systems can redistribute aspects of users' agency and autonomy \cite{chengConversationalAgentsYour2025, zhangExploringCollaborationPatterns2025}, which suggest a form of self-extension (cf., Clark's Extended Mind \cite{clarkExtendedMind1998b}), wherein technologies become tools (cf., \cite{bergstromToolExtensionHumanComputer2019, liAIShellUnderstanding2023a}) that are themselves part and parcel of the user's cognitive processes, reasoning patterns, and even vocal expressions. As such, while AI systems increasingly modify how users speak or express ideas through LLM-assisted writing \cite{Liang2025widespreadllms}, research typically focuses on users interacting with AI agents as external tools. There has been less attention on the experiential consequences of users’ own behavior becoming assimilated with intelligent output, forming a closed cognitive or sensorimotor loop. This latter perspective aligns more closely with visions of human–computer integration \cite{muellerNextStepsHumanComputer2020}, by which computational systems become interwoven with perception and action. Together, these developments invite exploring avatar-based self-extension, where AI-mediated modifications to speech and action are experienced as part of one’s own communicative behavior.

In this work, we present the early design and implementation of \textbf{ProxyMe}, a VR prototype that lets users embody a virtual avatar and experience how an AI system modifies the avatar's voice and spoken content on their behalf (\autoref{fig:teaser}). By combining avatar-based embodiment, voice modulation, and speech augmentation, ProxyMe seeks to explore the boundary between self-authored and AI-mediated communication. Conceptually, ProxyMe relates to multiple aspects of the JIZAI Body framework \cite{Inami2022jizai}, namely Fusionary, Duplicatory, and Possessory/Transformatory modes of being. Through this configuration of AI-augmented avatar embodiment (i.e., self-extension), ProxyMe enables the exploration of how such systems influence users' perceived agency, authorship, and self-identification with their avatar(s) and its actions. As external reasoning systems increasingly act on users' behalf in communication \cite{chengConversationalAgentsYour2025}, they do far more than supporting task completion and productivity (cf., \cite{Constantinides2025futureofwork}): When users reflect on their self-extended actions and speech, such systems can implicitly shape how users express opinions, make decisions, and form values. This is relevant as communicative actions are partially delegated to external systems where AI safety and human value alignment is paramount \cite{gyevnár2025aisafety,shen2025positionbidirectionalhumanaialignment}. At the same time, users may experience shifts in how they perceive authorship and agency \cite{Cornelio2022agencyhint,muellerNextStepsHumanComputer2020}. Our work envisions multiple scenarios where such novel interactions can benefit, including addressing speech impairment \cite{cappadonaVRChildLanguageDisorder2023}, public speaking \cite{bachmannVirtualRealityPublic2023}, mind-body therapy \cite{Dollinger2024bodysapping}, and self-understanding and (virtual) identity exploration \cite{Lorenzo2025VRselfunderstanding}. Understanding these experiential dynamics is therefore critical for designing AI-mediated communication systems that balance delegation with user agency, and can support responsible attribution of action and ownership.

\section{Related Work}
\label{sec:rel_work}
\subsection{Multimodal self-extension in VR} 

Clark's Extended Mind theory proposes that external resources such as notebooks and smartphones can be recognized as parts of mental processes, and thus as extensions of the biological brain, when they are integrated with cognitive tasks \cite{clarkExtendedMind1998b}. This theory has been examined in HCI research, especially in how such tool use can turn into measurable tool extensions \cite{bergstromToolExtensionHumanComputer2019}. And, regarding this examination, Virtual Reality (VR) provides a plausible context for examining this Extended Mind framework, where users embody and control their avatars. Furthermore, such extensions become particularly salient in immersive XR environments where they extend to the biological self \cite{bioccaCommunicationAgeVirtual2013, bioccaImmersiveVirtualReality1995}, where users can embody avatars and assimilate their behaviors (so-called Proteus effect \cite{yeeProteusEffectEffect2007, schwindTheseAreNot2017}), adopt alternative voices or identities \cite{fanMultiEmbodimentDigitalHumans2017,albrecht2025futureyoudesigningevaluating}, control multiple bodies \cite{miuraMultiSomaDistributedEmbodiment2021}, or even undergo real-time bodily transformations \cite{otonoTransformingEffectsVisual2023}. Additionally, when interacting with users in XR environments, AI agents can also gain awareness about context and give real-time feedback, enhancing users' sense of presence and engagement\cite{ruebConversingAIAgents2025, zhangPromptingEmbodiedAI2025, gunawardhanaUserAwareInteractiveVirtual2024}. Furthermore, voice modification can influence how we present ourselves and significantly affect self-perception and identity; Sounding like oneself in VR reinforces commitment and presence \cite{rahillEffectsAvatarPlayersimilarity2021}, while using a totally different voice can lead to discomfort in communication, reduced sense of agency, and a feeling of disconnection \cite{ohata_i_2022}. 
Indeed, self-cloned voice has also been used for coaching or nudging people toward a better self \cite{pataranutapornFutureYouConversation2024, fangLeveragingAIGeneratedEmotional2025}. Thus, based on these aspects, it is plausible to say that voice augmentation can effectively contribute to the extension of the self, however, it simultaneously raises the question: is it me?






\subsection{Agency and authorship in AI augmentation}

AI-driven assistance is increasingly integrated into daily workflows \cite{Liang2025widespreadllms}, where co-creating with such assistance signals a new era of cognitive self-extension, where human express meaning jointly with AI \cite{zhangExploringCollaborationPatterns2025, campbellExtendingSelfAImediated2025a}. Prior work has shown that users may appropriate AI suggestions as their own \cite{jakeschCoWritingOpinionatedLanguage2023}, that responsibility can diffuse in human-AI teams \cite{salatinoInfluenceAIBehavior2025,Brailsford2025responsibility}, and that such assistance can even create an AI memory gap where users mistake AI content as their own \cite{zindulka2025aimemorygapusers}. These findings motivate the concept of avatar self-extension, where system-generated expressions are experienced as part of the user’s own communicative behavior. We argue that this configuration gives rise to a distinct mode of AI-mediated, or more precisely, an AI-self-modified communication. In such situations, the boundaries of authorship and agency may become ambiguous (cf., \cite{liAIShellUnderstanding2023a,Mueller2021bodilyintegration}). Users may vary in whether they notice AI modifications (cf., perceptions of Human Augmentations \cite{Villa2023perceptionhumanaug}), how they attribute psychological ownership of expressed content (cf., \cite{Joshi2025ownership}), and whether they experience alignment or alienation from the resulting behavior (cf., human-AI alignment \cite{shen2025positionbidirectionalhumanaialignment}). Together, these works raise considerations for how to study human autonomy and agency under self-extension scenarios.

\section{ProxyMe Prototype}

We propose ProxyMe, a system designed to support avatar self-extension at both bodily and cognitive levels, through avatar-based embodiment and AI-assisted co-creation. We illustrate the envisioned interaction in \autoref{fig:teaser}. Below we describe the concept of ProxyMe, design goals, system implementation, and potential scenarios where it can impact user self-extension. Our current prototype interaction [video] shows a snippet moral scenario dialogue (cloned voice, argument modification) between the user's avatar self-extension and a virtual agent (see \purple{Supplementary Material} video).

\subsection{Design goals}


Based on this objective, and the considerations reported in Section 2, we derived the following design goals, where our current work focuses mainly on the first three:



\textbf{Acting on user's behalf.} Instead of ProxyMe requiring explicit user approval and guidance at every step, we investigate how partial delegation of communicative actions influences users' sense of authorship and agency. In our design, the virtual agent is allowed to perform modifications to the user's outputs, such as reformulating, or refining, based on user preferences, however still remaining attributable to the user. 

\textbf{Strong embodiment.} Strong embodiment is essential for self-extension, as it enables users to attribute AI-mediated actions to themselves rather than to an external system. ProxyMe is designed to provide embodied interaction through a virtual avatar in VR, where users can control the avatar's arms to establish visuomotor correspondence (cf., virtual body ownership \cite{Slater2009ownershipvirtualbodt} and self-avatars \cite{Boban2024embodiedselfavatar}).
Additionally, the use of the user's cloned voice reinforces continuity between the user and the avatar, which further strengthens user perceptions that ProxyMe actions originate from the self.

\textbf{Observable perspective.} 
Instead of enforcing continuous first-person embodiment, ProxyMe adopts an observable third-person perspective. This allows users to remain aware of ongoing AI-mediated actions without feeling obliged to constantly act or intervene. Prior work shows that third-person perspectives in VR can still support a sense of embodiment \cite{hoppe_there_2022}, which makes it possible to balance self-identification with the avatar and momentary disengagement.
\\

\noindent Below are broader design goals for building avatar self-extension systems, which are currently work in progress.

\textbf{Steerable.} Preserving user control is important despite partial delegation (cf., users seek to maintain a sense of agency with autonomous conversational systems \cite{chengConversationalAgentsYour2025}). In future iterations, prompting strategies and user preference selection could enable user dynamic control over AI-mediated speech. However, in our present prototype design, speech mediation is constrained to predefined conditions, to better allow for controlled user testing.

\textbf{Real-time interaction and low latency.}
Self-extension will depend on a tight temporal coupling between intention, expression, and the experienced feedback. Excessive delays risk disrupting attribution and shifting the experience from self-extension to an external automation. This is an ongoing task to ensure a seamless user experience, where our key mitigation mechanism is to draw on perceptual masking during the experiment.

\textbf{Privacy-preserving and local execution.} 
Given ProxyMe operates on users’ voice, speech content, and embodied behavior, privacy preservation is a core design requirement. The system is designed to run locally, avoiding the transmission or external storage of any sensitive biometric and communicative data. 

\textbf{Human value alignment.} In our current design, ProxyMe prioritizes conservative, user-aligned transformations and to avoid introducing content that diverges from the user’s expressed stance in a harmful manner. We treat value alignment here as both a technical AI safety objective and as an experiential concern for studying AI-mediated communication.

\subsection{System implementation}
\begin{figure*}[t]
	\centering
	\subfigure[ProxyMe takes user's initial utterance, transcribes it into text, makes textual modification, and finally synthesizes the modified speech with voice cloning, which is played as avatar's second round output in dialogues.]{\label{fig:system}\includegraphics[width=0.49\linewidth]
    {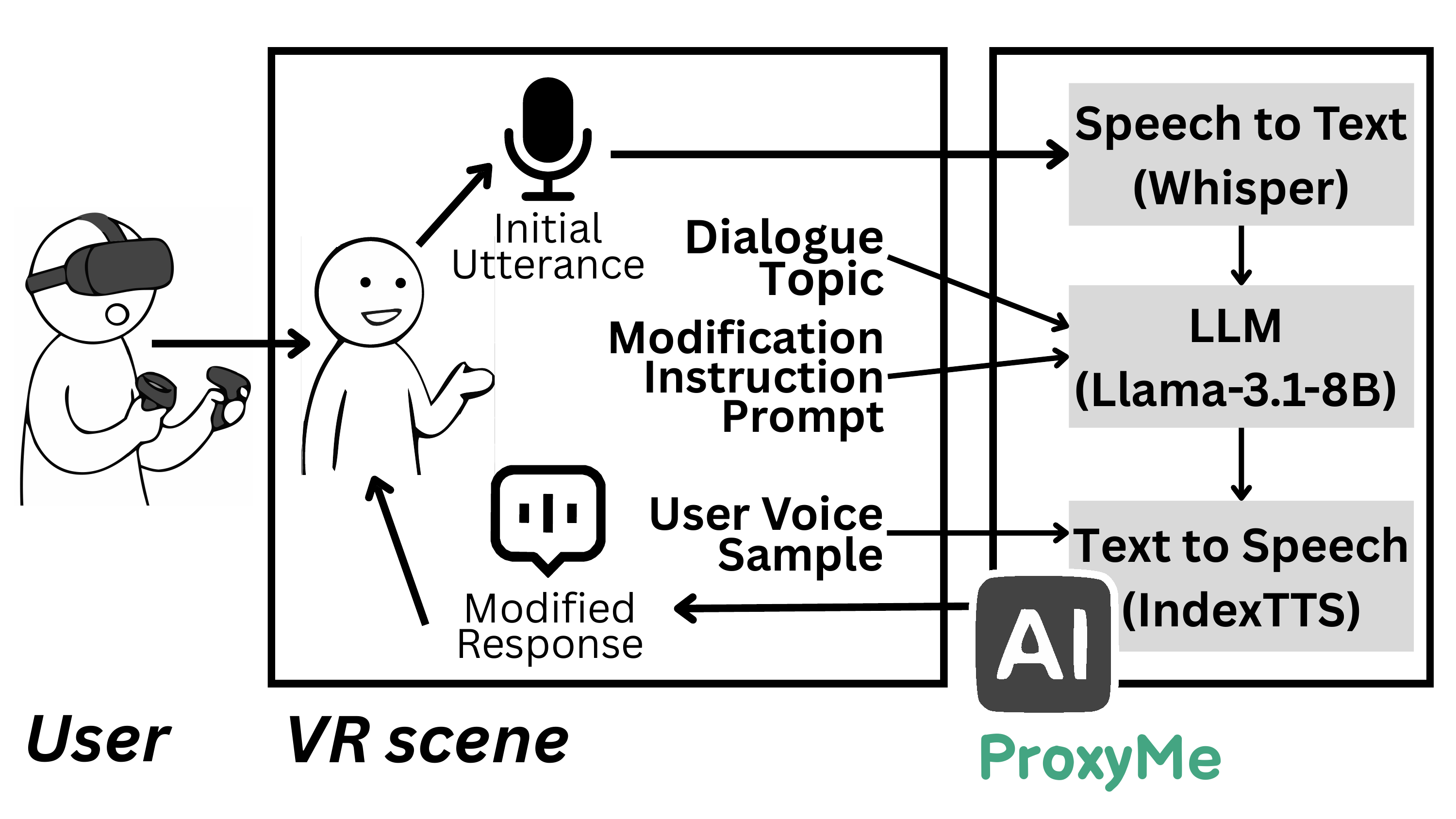}}\Description{A schematic diagram of the ProxyMe system pipeline within a virtual reality (VR) interaction. On the left, a user wearing a VR headset and holding controllers interacts with a virtual agent inside a box labeled “VR scene.” The agent produces an “Initial Utterance” (microphone icon), which enters a “Dialogue Topic” module.

    Within the VR scene box, the user’s spoken input is labeled “User Voice Sample,” and the agent’s reply is labeled “Modified Response.” A “Modification Instruction Prompt” influences how the dialogue is altered.

    Arrows lead from the VR scene to a processing pipeline on the right. The user’s speech is first converted using “Speech to Text (Whisper),” then processed by a large language model labeled “LLM (Llama-3.1-8B),” and finally synthesized back into audio through “Text to Speech (IndexTTS).” These components are grouped under the label “AI ProxyMe.”

    The diagram illustrates how user speech in VR is transcribed, modified by a language model according to an instruction prompt, and re-synthesized into spoken output before being delivered back into the virtual interaction.} \hfill
	\subfigure[User (left) controlling their proxy that converses with an agent in VR (right).]{\label{fig:vrsetup}\includegraphics[width=0.49\linewidth]{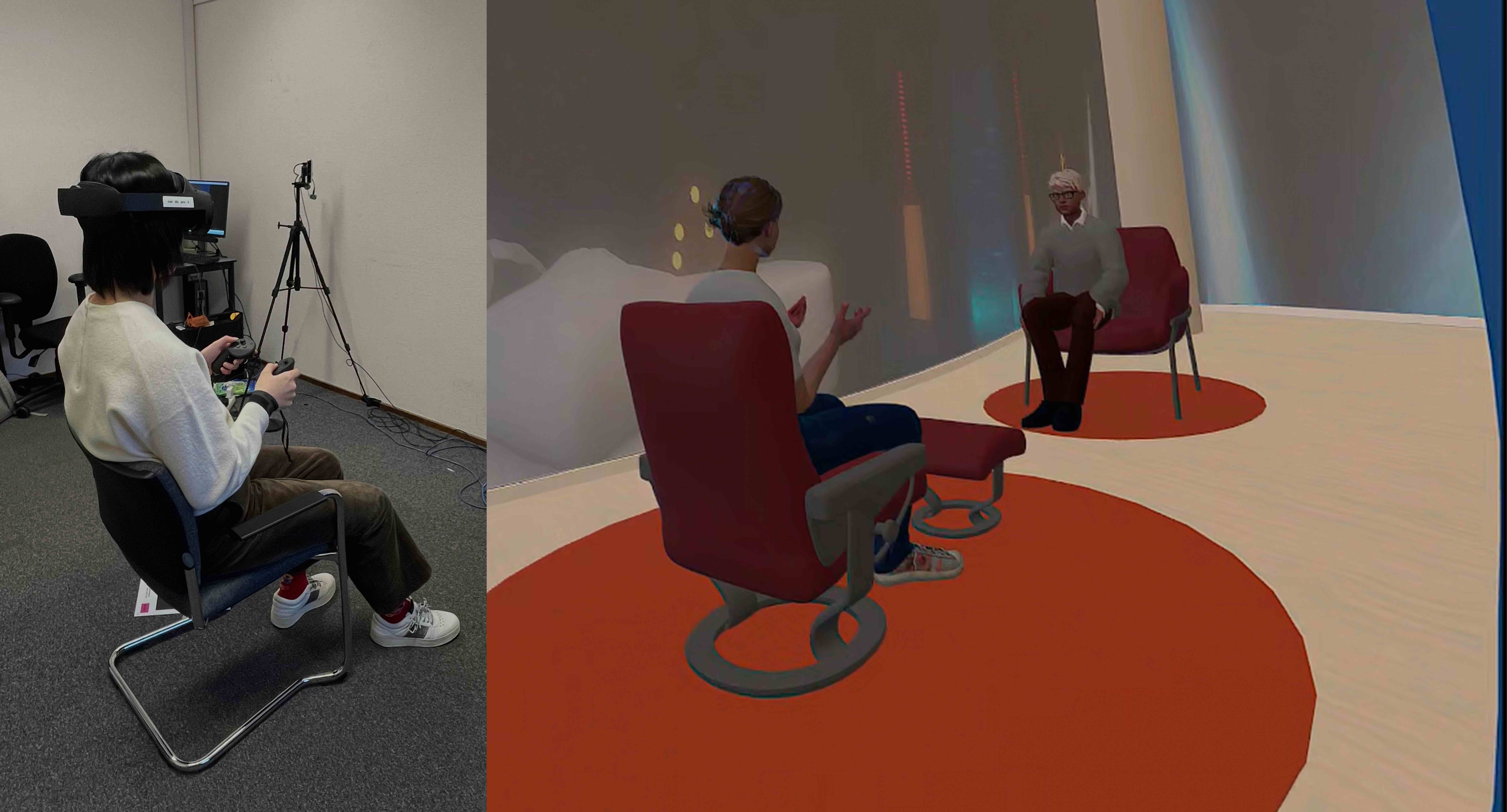}}\Description{A photograph showing a mixed physical and virtual reality setup. On the left, a seated user in a laboratory room wears a VR headset and holds handheld controllers. A tripod and computer equipment are visible in the background.

On the right side of the image, the virtual reality view is shown: the user’s avatar sits in a red armchair facing another seated avatar in a matching chair across a small round table. The two avatars appear to be engaged in conversation in a minimalist virtual room with light-colored walls and a circular rug beneath the chairs.

The image illustrates a user (left) physically controlling their proxy avatar, which converses with an agent inside the VR environment (right).}
	\caption{System overview and experimental setup of ProxyMe. }
	
	\label{fig:finalprototype}
\end{figure*}

Our current system overview is shown in Fig. \ref{fig:system}. The prototype is implemented in Unity, where users are embodied as an avatar interacting with virtual agents or other participants, where they have a dialogue, initiated by the agent ask a question (Fig. \ref{fig:vrsetup}). Avatars were created using ReadyPlayerMe\footnote{\url{https://readyplayer.me}} and Mixamo\footnote{\url{https://www.mixamo.com/}}. ProxyMe takes the user's initial vocal speech, transcribes it using the Whisper model\footnote{\url{https://openai.com/index/whisper/}}), where the text is then fed into Llama-3.1-8B\footnote{\url{https://huggingface.co/meta-llama/Llama-3.1-8B}}), which edits, extends, or reformulates content based on predefined user prompts. To render the output into following up avatar vocal answers (i.e., text to speech (TTS)), we use the IndexTTS\footnote{\url{https://github.com/index-tts/index-tts}} model, to generate either a cloned version of the user's voice or a neutral (Siri-like) synthetic voice. The agent uses a predefined voice from ElevenLabs\footnote{\url{https://elevenlabs.io}}, varied according to gender.






\subsection{Envisioned scenarios}
We illustrate the use of ProxyMe through three scenarios that emphasize assistive communication, identity exploration, and therapeutic distancing via mediated self-expression. \textbf{Public speaking and assistive support;} \textit{Context \& User Goal:} The user struggles with clarity, fluency, or expressive tone during public speaking or high-pressure communication. The goal is to retain authorship over spoken content while improving delivery (e.g., confidence, empathy, fluency)~\cite{huangAIVRSpeakAnxiety2026, Suranga2024assistive}. \textit{Mediated Interaction:} As the user speaks naturally, ProxyMe re-expresses the speech through a virtual proxy with minimal, real-time modifications (e.g., reduced disfluencies, adjusted prosody), preserving semantic intent while improving perceived delivery. \textbf{Role-playing and virtual identity exploration;} \textit{Context \& User Goal:} The user aims to explore alternative expressive styles or partial identities (e.g., more assertive or empathetic) while preserving personal values and intent~\cite{Lorenzo2025VRselfunderstanding}. \textit{Mediated Interaction: } ProxyMe mediates the user’s speech to emphasize selected expressive traits, enabling immediate experience of different self-presentations during interaction without adopting a fully fictional character~\cite{pataranutapornFutureYouConversation2024}. \textbf{Mind-body therapy;} \textit{Context \& User Goal:} The user seeks psychological distance from emotionally charged speech for reflective or therapeutic purposes~\cite{Dollinger2024bodysapping}. \textit{Mediated Interaction: } ProxyMe externalizes the user’s speech through a virtual proxy or alternative embodiment, enabling the user to observe and respond to their own statements with reduced emotional load.

\section{Next steps for developing and evaluating ProxyMe}

First, our early prototype is implemented as a constrained, single-round interaction, leaving open whether ProxyMe should integrate long-term user preferences and accumulate memory. Second, measured end-to-end latency from speech input to AI-mediated output currently sits at approximately 11.6 sec (stt: 1.2, llm: 2.9, tts: 7.5 across 200 runs), which may affect interaction flow and sense of embodiment and agency; however, we are currently exploring means to reduce subjective latency through embedding processing time with conversational turn-taking window (i.e., perceptual masking). We will also carry out pilot studies to examine the extent these delays impact user experience. We will furthermore explore how to make our TTS model faster, given it is responsible for much of the latency. We will consider transferring from batch synthesis to a streaming architecture, allowing for chunk-level buffering for immediate playback, significantly reducing perceived latency. Third, we are aware ethical concerns may arise regarding user acceptance and potential alienation of hearing one's own (cloned) voice, yet perceptibly different from their real voice \cite{dielDeviationTypicalOrganic2024a}. Similarly, one may possibly experience confusion when listening to one's arguments being reformulated, challenged, or reversed. Moreover, a single user might differ across contexts how they prioritize different principles, which influences their tolerance for delegating control to their self-extensions. 


We anticipate running a 2 (Voice: cloned vs. robotic) × 3 (Content: repetition, enhancement, countered conclusion) within-subjects design for testing ProxyMe in a controlled user study (cf., Fig.\ref{fig:teaser}). Participants engage in short turn-based VR moral dialogues (drawn from MoralChoice dataset \cite{scherrerEvaluatingMoralBeliefs2023}) with a virtual LLM-powered agent in which their initial utterance is followed by an AI-mediated continuation (self-extension). After each trial, participants assess using self-reports their perceived agency and authorship \cite{salatinoInfluenceAIBehavior2025}. This allows us to examine how different degrees of delegation and voice modulation affect their self-extension experiences and overall UX.
We will pilot test how users experience their self-extension, and how this can be integrated into the interaction loop through mechanisms such as pausing or restarting the ProxyMe output, as well as dynamically adjusting levels of autonomy on demand (cf. \cite{Collier2025sharedautonomy}). 
We also consider extensions of ProxyMe to mixed reality environments (e.g., AR) as future work. Understanding how such variations in delegation across immersive interactions influence agency, attribution, and self-identification will help us develop both an ethically sensitive and practically viable experimental setup.

\section{Considerations for designing avatar self-extensions in VR}

A key question that arises concerns social attribution: how users take responsibility for actions performed by avatar extensions as belonging to the self, especially when users disagree with AI-mediated output. This raises the question to the extent we should discourage such AI extensions in the first place (cf., \cite{naeemShouldWeDiscourage2024}). While attributing undesirable outcomes to the AI system, easily claiming "It wasn't me!" may serve as a short-term strategy, longer-term interactions raise unresolved questions about responsibility and authorship as AI-mediated expression is integrated into everyday tasks (cf., \cite{Brailsford2025responsibility,Joshi2025ownership,Liang2025widespreadllms}). Prior research shows that users already tend to misattribute language model outputs to their own beliefs and intentions \cite{jakeschCoWritingOpinionatedLanguage2023}. With persistent avatar extensions, users may gradually lose track (whether phenomenologically or cognitively) of which content originated from themselves versus from the system (cf., AI memory gap \cite{zindulka2025aimemorygapusers}). As such, it may be necessary for future systems to support mechanisms for distinguishing and managing multiple sources of expression (self vs extension), for example by explicitly tracking the provenance of actions and speech across different origins, potentially through separate or labeled memory representations (cf., Ergo Proxy). 

The use of voice and avatar cloning introduces risks of identity misuse and appropriation \cite{Sobolev2025authenticity}, which raises concerns about virtual identity theft and human digital doubles. Furthermore, as future systems increasingly learn from contexts and accumulate interaction history (e.g., boosting model memory with RAG-based techniques \cite{Qian2025memoryrag}), avatar extensions may shift from single-use conversational tools towards persistent social actors that form sustained social relationships on the user's behalf (cf., strong avatar-as-proxy relations \cite{sweeneyAvatarsProxies2023}). This raises questions about how others relate to these extensions, even if their artificial nature is transparently disclosed \cite{Elali2024aidisclosure}. Are they treated as equivalent to, distinct from, or even preferable to the user’s unmediated self. Together, these dynamics may further destabilize the boundary between self and proxy, presenting a critical challenge for HCI research.

\section{AI Disclosure Usage}
We used GPT-5.2 to sometimes suggest clarity and structure in writing, and Gemini Nano Banana Pro to co-create Fig 1. Otherwise, conceptualization, analytical decisions, and verification were human-only.








\bibliographystyle{ACM-Reference-Format}
\interlinepenalty=10000
\bibliography{bib_extSelfVR}

@article{sweeneyAvatarsProxies2023,
  title = {Avatars as {{Proxies}}},
  author = {Sweeney, Paula},
  date = {2023-09},
  JOURNAL = {Minds and Machines},
  shortjournal = {Minds \& Machines},
  volume = {33},
  number = {3},
  year = {2023}, 
  pages = {525--539},
  issn = {0924-6495, 1572-8641},
  doi = {10.1007/s11023-023-09643-z},
  url = {https://link.springer.com/10.1007/s11023-023-09643-z},
  urldate = {2026-03-02},
  langid = {english}
}

@ARTICLE{Slater2009ownershipvirtualbodt,
    
AUTHOR={Slater, Mel  and PÃ©rez Marcos, Daniel  and Ehrsson, Henrik  and Sanchez-Vives, Maria V.},
           
TITLE={Inducing illusory ownership of a virtual body},
          
JOURNAL={Frontiers in Neuroscience},
          
VOLUME={Volume 3 - 2009},
  
YEAR={2009},
  
URL={https://www.frontiersin.org/journals/neuroscience/articles/10.3389/neuro.01.029.2009},
  
DOI={10.3389/neuro.01.029.2009},
  
ISSN={1662-453X}}

@ARTICLE{Boban2024embodiedselfavatar,
  author={Boban, Loën and Boulic, Ronan and Herbelin, Bruno},
  journal={IEEE Transactions on Visualization and Computer Graphics}, 
  title={In Case of Doubt, One Follows One's Self: The Implicit Guidance of the Embodied Self-Avatar}, 
  year={2024},
  volume={30},
  number={5},
  pages={2109-2118},
  doi={10.1109/TVCG.2024.3372042}}

@inproceedings{Suranga2024assistive,
author = {Nanayakkara, Suranga Chandima and Inami, Masahiko and Mueller, Florian and Huber, Jochen and Gupta, Chitralekha and Jouffrais, Christophe and Kunze, Kai and Patibanda, Rakesh and Chan, Samantha W T and Messerschmidt, Moritz Alexander},
title = {Exploring the Design Space of Assistive Augmentation},
year = {2023},
isbn = {9781450399845},
publisher = {Association for Computing Machinery},
address = {New York, NY, USA},
url = {https://doi.org/10.1145/3582700.3582729},
doi = {10.1145/3582700.3582729},
booktitle = {Proceedings of the Augmented Humans International Conference 2023},
pages = {371–373},
numpages = {3},
location = {Glasgow, United Kingdom},
series = {AHs '23}
}

@article{huangAIVRSpeakAnxiety2026,
  title = {The Effectiveness of an {{AI-integrated VR}} Oral Training Application in Reducing Public Speaking Anxiety and Interview Anxiety},
  author = {Huang, Peiwen and Hwang, Yanling and Hsu, Jui Ling and Peng, Chien Fand and Tsai, Cheng Han and Wang, Chih Yao},
  year = 2026,
  month = jun,
  journal = {Computers and Education: Artificial Intelligence},
  volume = {10},
  pages = {100514},
  issn = {2666920X},
  doi = {10.1016/j.caeai.2025.100514},
  urldate = {2026-01-22},
  langid = {english}
}

@article{bachmannVirtualRealityPublic2023,
  title = {Virtual Reality Public Speaking Training: Effectiveness and User Technology Acceptance},
  shorttitle = {Virtual Reality Public Speaking Training},
  author = {Bachmann, Manuel and Subramaniam, Abimanju and Born, Jonas and Weibel, David},
  year = 2023,
  month = sep,
  journal = {Frontiers in Virtual Reality},
  volume = {4},
  pages = {1242544},
  issn = {2673-4192},
  doi = {10.3389/frvir.2023.1242544},
  urldate = {2026-01-22}
}

@article{cappadonaVRChildLanguageDisorder2023,
  title = {Feasibility and {{Effectiveness}} of {{Speech Intervention Implemented}} with a {{Virtual Reality System}} in {{Children}} with {{Developmental Language Disorders}}: {{A Pilot Randomized Control Trial}}},
  shorttitle = {Feasibility and {{Effectiveness}} of {{Speech Intervention Implemented}} with a {{Virtual Reality System}} in {{Children}} with {{Developmental Language Disorders}}},
  author = {Cappadona, Irene and Ielo, Augusto and La Fauci, Margherita and Tresoldi, Maria and Settimo, Carmela and De Cola, Maria Cristina and Muratore, Rosalia and De Domenico, Carmela and Di Cara, Marcella and Corallo, Francesco and Tripodi, Emanuela and Impallomeni, Caterina and Quartarone, Angelo and Cucinotta, Francesca},
  year = 2023,
  month = aug,
  journal = {Children},
  volume = {10},
  number = {8},
  pages = {1336},
  issn = {2227-9067},
  doi = {10.3390/children10081336},
  urldate = {2026-01-22},
  langid = {english}
}

@article{Lorenzo2025VRselfunderstanding,
author = {Lorenzo Antichi and Lorenzo BaglÃ¬o and Chiara Rossi and Giuseppe Riva},
title ={Introspecta VR: The Use of Virtual Reality and Artificial Intelligence for Self-Understanding, Future Self-Identification, and Personal Transformation},

journal = {Cyberpsychology, Behavior, and Social Networking},
volume = {28},
number = {6},
pages = {447-449},
year = {2025},
doi = {10.1089/cyber.2025.0172},
    note ={PMID: 40391453},

URL = { 
    
    
        https://journals.sagepub.com/doi/abs/10.1089/cyber.2025.0172
    

},
eprint = { 
    
    
        https://journals.sagepub.com/doi/pdf/10.1089/cyber.2025.0172
    

}

}

@misc{wikighostintheshell,
  title        = {Ghost in the Shell},
  howpublished = {\url{https://en.wikipedia.org/wiki/Ghost_in_the_Shell}},
  note         = {Wikipedia, The Free Encyclopedia. Accessed: 2026-01-22},
  year         = {2026},
  publisher    = {Wikimedia Foundation}
}

@misc{wikiergoproxy,
  title        = {Ergo Proxy},
  howpublished = {\url{https://en.wikipedia.org/wiki/Ergo_Proxy}},
  note         = {Wikipedia, The Free Encyclopedia. Accessed: 2026-01-22},
  year         = {2026},
  publisher    = {Wikimedia Foundation}
}

@misc{zindulka2025aimemorygapusers,
      title={The AI Memory Gap: Users Misremember What They Created With AI or Without}, 
      author={Tim Zindulka and Sven Goller and Daniela Fernandes and Robin Welsch and Daniel Buschek},
      year={2025},
      eprint={2509.11851},
      archivePrefix={arXiv},
      primaryClass={cs.HC},
      url={https://arxiv.org/abs/2509.11851}, 
}

@inproceedings{Collier2025sharedautonomy,
author = {Collier, Maggie A and Narayan, Rithika and Admoni, Henny},
title = {The Sense of Agency in Assistive Robotics Using Shared Autonomy},
year = {2025},
publisher = {IEEE Press},
booktitle = {Proceedings of the 2025 ACM/IEEE International Conference on Human-Robot Interaction},
pages = {880–888},
numpages = {9},
location = {Melbourne, Australia},
series = {HRI '25}
}

@inproceedings{Qian2025memoryrag,
author = {Qian, Hongjin and Liu, Zheng and Zhang, Peitian and Mao, Kelong and Lian, Defu and Dou, Zhicheng and Huang, Tiejun},
title = {MemoRAG: Boosting Long Context Processing with Global Memory-Enhanced Retrieval Augmentation},
year = {2025},
isbn = {9798400712746},
publisher = {Association for Computing Machinery},
address = {New York, NY, USA},
url = {https://doi.org/10.1145/3696410.3714805},
doi = {10.1145/3696410.3714805},
booktitle = {Proceedings of the ACM on Web Conference 2025},
pages = {2366–2377},
numpages = {12},
location = {Sydney NSW, Australia},
series = {WWW '25}
}

@article{Sobolev2025authenticity,
author = {Anton Sobolev},
title = {The last call for authenticity: AI reshaping voice fraud landscape},
journal = {Journal of Cyber Policy},
volume = {0},
number = {0},
pages = {1--21},
year = {2025},
publisher = {Routledge},
doi = {10.1080/23738871.2025.2597191},


URL = { 
    
        https://doi.org/10.1080/23738871.2025.2597191
    
    

},
eprint = { 
    
        https://doi.org/10.1080/23738871.2025.2597191
    
    

}

}

@inproceedings{Dollinger2024bodysapping,
author = {D\"{o}llinger, Nina and Mal, David and Keppler, Sebastian and Wolf, Erik and Botsch, Mario and Israel, Johann Habakuk and Latoschik, Marc Erich and Wienrich, Carolin},
title = {Virtual Body Swapping: A VR-Based Approach to Embodied Third-Person Self-Processing in Mind-Body Therapy},
year = {2024},
isbn = {9798400703300},
publisher = {Association for Computing Machinery},
address = {New York, NY, USA},
url = {https://doi.org/10.1145/3613904.3642328},
doi = {10.1145/3613904.3642328},
booktitle = {Proceedings of the 2024 CHI Conference on Human Factors in Computing Systems},
articleno = {110},
numpages = {18},
location = {Honolulu, HI, USA},
series = {CHI '24}
}

@article{gyevnár2025aisafety,
  title   = {AI safety for everyone},
  author  = {Gyevn{\'a}r, Bence and Kasirzadeh, Atoosa},
  journal = {Nature Machine Intelligence},
  volume  = {7},
  pages   = {531--542},
  year    = {2025},
  doi     = {10.1038/s42256-025-01020-y},
  url     = {https://doi.org/10.1038/s42256-025-01020-y}
}

@misc{albrecht2025futureyoudesigningevaluating,
      title={Future You: Designing and Evaluating Multimodal AI-generated Digital Twins for Strengthening Future Self-Continuity}, 
      author={Constanze Albrecht and Chayapatr Archiwaranguprok and Rachel Poonsiriwong and Awu Chen and Peggy Yin and Monchai Lertsutthiwong and Kavin Winson and Hal Hershfield and Pattie Maes and Pat Pataranutaporn},
      year={2025},
      eprint={2512.06106},
      archivePrefix={arXiv},
      primaryClass={cs.HC},
      url={https://arxiv.org/abs/2512.06106}, 
}

@inproceedings{Inami2022jizai,
author = {Inami, Masahiko and Uriu, Daisuke and Kashino, Zendai and Yoshida, Shigeo and Saito, Hiroto and Maekawa, Azumi and Kitazaki, Michiteru},
title = {Cyborgs, Human Augmentation, Cybernetics, and JIZAI Body},
year = {2022},
isbn = {9781450396325},
publisher = {Association for Computing Machinery},
address = {New York, NY, USA},
url = {https://doi.org/10.1145/3519391.3519401},
doi = {10.1145/3519391.3519401},
booktitle = {Proceedings of the Augmented Humans International Conference 2022},
pages = {230–242},
numpages = {13},
location = {Kashiwa, Chiba, Japan},
series = {AHs '22}
}

@ARTICLE{Cornelio2022agencyhint,
    
AUTHOR={Cornelio, Patricia  and Haggard, Patrick  and Hornbaek, Kasper  and Georgiou, Orestis  and BergstrÃ¶m, Joanna  and Subramanian, Sriram  and Obrist, Marianna },
           
TITLE={The sense of agency in emerging technologies for human computer integration: A review},
          
JOURNAL={Frontiers in Neuroscience},
          
VOLUME={Volume 16 - 2022},
  
YEAR={2022},
  
URL={https://www.frontiersin.org/journals/neuroscience/articles/10.3389/fnins.2022.949138},
  
DOI={10.3389/fnins.2022.949138},
  
ISSN={1662-453X}}

@inproceedings{Elali2024aidisclosure,
author = {El Ali, Abdallah and Venkatraj, Karthikeya Puttur and Morosoli, Sophie and Naudts, Laurens and Helberger, Natali and Cesar, Pablo},
title = {Transparent AI Disclosure Obligations: Who, What, When, Where, Why, How},
year = {2024},
isbn = {9798400703317},
publisher = {Association for Computing Machinery},
address = {New York, NY, USA},
url = {https://doi.org/10.1145/3613905.3650750},
doi = {10.1145/3613905.3650750},
booktitle = {Extended Abstracts of the CHI Conference on Human Factors in Computing Systems},
articleno = {342},
numpages = {11},
location = {Honolulu, HI, USA},
series = {CHI EA '24}
}

@inproceedings{Brailsford2025responsibility,
author = {Brailsford, Joe and Vetere, Frank and Velloso, Eduardo},
title = {Responsibility Attribution in Human Interactions with Everyday AI Systems},
year = {2025},
isbn = {9798400713941},
publisher = {Association for Computing Machinery},
address = {New York, NY, USA},
url = {https://doi.org/10.1145/3706598.3713126},
doi = {10.1145/3706598.3713126},
booktitle = {Proceedings of the 2025 CHI Conference on Human Factors in Computing Systems},
articleno = {1020},
numpages = {17},
location = {
},
series = {CHI '25}
}

@inproceedings{Constantinides2025futureofwork,
author = {Constantinides, Marios and Verma, Himanshu and Sadeghian, Shadan and El Ali, Abdallah},
title = {The Future of Work is Blended, Not Hybrid},
year = {2025},
isbn = {9798400713842},
publisher = {Association for Computing Machinery},
address = {New York, NY, USA},
url = {https://doi.org/10.1145/3729176.3729202},
doi = {10.1145/3729176.3729202},
booktitle = {Proceedings of the 4th Annual Symposium on Human-Computer Interaction for Work},
articleno = {28},
numpages = {13},
location = {
},
series = {CHIWORK '25}
}

@misc{shen2025positionbidirectionalhumanaialignment,
      title={Position: Towards Bidirectional Human-AI Alignment}, 
      author={Hua Shen and Tiffany Knearem and Reshmi Ghosh and Kenan Alkiek and Kundan Krishna and Yachuan Liu and Ziqiao Ma and Savvas Petridis and Yi-Hao Peng and Li Qiwei and Sushrita Rakshit and Chenglei Si and Yutong Xie and Jeffrey P. Bigham and Frank Bentley and Joyce Chai and Zachary Lipton and Qiaozhu Mei and Rada Mihalcea and Michael Terry and Diyi Yang and Meredith Ringel Morris and Paul Resnick and David Jurgens},
      year={2025},
      eprint={2406.09264},
      archivePrefix={arXiv},
      primaryClass={cs.HC},
      url={https://arxiv.org/abs/2406.09264}, 
}

@inproceedings{Joshi2025ownership,
author = {Joshi, Nikhita and Vogel, Daniel},
title = {Writing with AI Lowers Psychological Ownership, but Longer Prompts Can Help},
year = {2025},
isbn = {9798400715273},
publisher = {Association for Computing Machinery},
address = {New York, NY, USA},
url = {https://doi.org/10.1145/3719160.3736608},
doi = {10.1145/3719160.3736608},
booktitle = {Proceedings of the 7th ACM Conference on Conversational User Interfaces},
articleno = {72},
numpages = {17},
location = {
},
series = {CUI '25}
}

@inproceedings{Villa2023perceptionhumanaug,
author = {Villa, Steeven and Niess, Jasmin and Nakao, Takuro and Lazar, Jonathan and Schmidt, Albrecht and Machulla, Tonja-Katrin},
title = {Understanding Perception of Human Augmentation: A Mixed-Method Study},
year = {2023},
isbn = {9781450394215},
publisher = {Association for Computing Machinery},
address = {New York, NY, USA},
url = {https://doi.org/10.1145/3544548.3581485},
doi = {10.1145/3544548.3581485},
booktitle = {Proceedings of the 2023 CHI Conference on Human Factors in Computing Systems},
articleno = {65},
numpages = {16},
location = {Hamburg, Germany},
series = {CHI '23}
}

@article{Mueller2021bodilyintegration,
title = {Towards understanding the design of bodily integration},
journal = {International Journal of Human-Computer Studies},
volume = {152},
pages = {102643},
year = {2021},
issn = {1071-5819},
doi = {https://doi.org/10.1016/j.ijhcs.2021.102643},
url = {https://www.sciencedirect.com/science/article/pii/S1071581921000616},
author = {Florian â Floydâ Mueller and Pedro Lopes and Josh Andres and Richard Byrne and Nathan Semertzidis and Zhuying Li and Jarrod Knibbe and Stefan Greuter}
}

@inproceedings{bergstromToolExtensionHumanComputer2019,
  title = {Tool {{Extension}} in {{Human-Computer Interaction}}},
  booktitle = {Proceedings of the 2019 {{CHI Conference}} on {{Human Factors}} in {{Computing Systems}}},
  author = {Bergstr{\"o}m, Joanna and Mottelson, Aske and Muresan, Andreea and Hornb{\ae}k, Kasper},
  year = 2019,
  month = may,
  pages = {1--11},
  publisher = {ACM},
  address = {Glasgow Scotland Uk},
  doi = {10.1145/3290605.3300798},
  urldate = {2026-01-21},
  isbn = {978-1-4503-5970-2},
  langid = {english}
}

@article{ruebConversingAIAgents2025,
  title = {Conversing with {{AI}} Agents in {{VR}}: {{An}} Early Investigation of Alignment and Modality},
  shorttitle = {Conversing with {{AI}} Agents in {{VR}}},
  author = {Rueb, Frederik and Sra, Misha},
  year = 2025,
  journal = {Empathic Computing},
  doi = {10.70401/ec.2025.0013},
  urldate = {2026-01-21},
  langid = {english}
}

@inproceedings{gunawardhanaUserAwareInteractiveVirtual2024,
  title = {Toward {{User-Aware Interactive Virtual Agents}}: {{Generative Multi-Modal Agent Behaviors}} in {{VR}}},
  shorttitle = {Toward {{User-Aware Interactive Virtual Agents}}},
  booktitle = {2024 {{IEEE International Symposium}} on {{Mixed}} and {{Augmented Reality}} ({{ISMAR}})},
  author = {Gunawardhana, Bhasura S. and Zhang, Yunxiang and Sun, Qi and Deng, Zhigang},
  year = 2024,
  month = oct,
  pages = {1068--1077},
  publisher = {IEEE},
  address = {Bellevue, WA, USA},
  doi = {10.1109/ISMAR62088.2024.00123},
  urldate = {2026-01-21},
  copyright = {https://doi.org/10.15223/policy-029},
  isbn = {979-8-3315-1647-5},
  langid = {english}
}

@inproceedings{zhangPromptingEmbodiedAI2025,
  title = {Prompting an {{Embodied AI Agent}}: {{How Embodiment}} and {{Multimodal Signaling Affects Prompting Behaviour}}},
  shorttitle = {Prompting an {{Embodied AI Agent}}},
  booktitle = {Proceedings of the 2025 {{CHI Conference}} on {{Human Factors}} in {{Computing Systems}}},
  author = {Zhang, Tianyi and Au Yeung, Colin and Aurelia, Emily and Onishi, Yuki and Chulpongsatorn, Neil and Li, Jiannan and Tang, Anthony},
  year = 2025,
  month = apr,
  pages = {1--25},
  publisher = {ACM},
  address = {Yokohama Japan},
  doi = {10.1145/3706598.3713110},
  urldate = {2026-01-21},
  isbn = {979-8-4007-1394-1},
  langid = {english}
}

@book{bioccaCommunicationAgeVirtual2013,
  title = {Communication in the Age of Virtual Reality},
  editor = {Biocca, Frank and Levy, Mark R.},
  year = 2013,
  month = feb,
  publisher = {Routledge},
  address = {New York},
  doi = {10.4324/9781410603128},
  isbn = {978-1-4106-0312-8}
}

@incollection{bioccaImmersiveVirtualReality1995,
  title = {Immersive Virtual Reality Technology},
  booktitle = {Communication in the Age of Virtual Reality},
  author = {Biocca, Frank and Delaney, Ben},
  year = 1995,
  month = jun,
  pages = {57--124},
  publisher = {L. Erlbaum Associates Inc.},
  address = {USA},
  urldate = {2025-07-24},
  isbn = {978-0-8058-1550-4}
}

@article{campbellExtendingSelfAImediated2025a,
  title = {Extending the Self through {{AI-mediated}} Communication: Functional, Ontological, and Anthropomorphic Extensions},
  shorttitle = {Extending the Self through {{AI-mediated}} Communication},
  author = {Campbell, Scott and Ellison, Nicole and Ross, Morgan},
  year = 2025,
  month = dec,
  journal = {Communication and Change},
  volume = {1},
  pages = {1--16},
  doi = {10.1007/s44382-025-00003-2}
}

@inproceedings{chengConversationalAgentsYour2025,
  title = {Conversational Agents on Your Behalf: {{Opportunities}} and Challenges of Shared Autonomy in Voice Communication for Multitasking},
  shorttitle = {Conversational Agents on Your Behalf},
  booktitle = {Proceedings of the 2025 {{CHI}} Conference on Human Factors in Computing Systems},
  author = {Cheng, Yi Fei and Shirado, Hirokazu and Kasahara, Shunichi},
  year = 2025,
  month = apr,
  pages = {1--18},
  publisher = {ACM},
  address = {Yokohama Japan},
  doi = {10.1145/3706598.3714017},
  urldate = {2025-07-08},
  eventtitle = {{{CHI}} 2025: {{CHI}} Conference on Human Factors in Computing Systems},
  langid = {english}
}

@article{chernyakovAIEthicsWhen2022,
  title = {{{AI Ethics}}: {{When}} Does an {{AI}} Voice Agent Need to Disclose Itself as an {{AI}} Agent?},
  shorttitle = {{{AI}} Ethics},
  author = {Chernyakov, Michael and Spriestersbach, Kai},
  year = 2022,
  doi = {10.13140/RG.2.2.26931.71205},
  urldate = {2025-05-21},
  langid = {english}
}

@article{clarkExtendedMind1998b,
  title = {The Extended Mind},
  author = {Clark, Andy and Chalmers, David},
  year = 1998,
  journal = {Analysis},
  volume = {58},
  number = {1},
  eprint = {3328150},
  eprinttype = {jstor},
  pages = {7--19},
  issn = {0003-2638},
  urldate = {2025-07-21}
}

@inproceedings{danryWearableReasonerEnhanced2020,
  title = {Wearable Reasoner: {{Towards}} Enhanced Human Rationality through a Wearable Device with an Explainable {{AI}} Assistant},
  shorttitle = {Wearable Reasoner},
  booktitle = {Proceedings of the Augmented Humans International Conference},
  author = {Danry, Valdemar and Pataranutaporn, Pat and Mao, Yaoli and Maes, Pattie},
  year = 2020,
  month = mar,
  pages = {1--12},
  publisher = {ACM},
  address = {Kaiserslautern Germany},
  doi = {10.1145/3384657.3384799},
  urldate = {2025-07-10},
  eventtitle = {{{AHs}} '20: {{Augmented}} Humans International Conference},
  isbn = {978-1-4503-7603-7},
  langid = {english}
}

@article{dielDeviationTypicalOrganic2024a,
  title = {Deviation from Typical Organic Voices Best Explains a Vocal Uncanny Valley},
  author = {Diel, Alexander and Lewis, Michael},
  year = 2024,
  month = may,
  journal = {Computers in Human Behavior Reports},
  volume = {14},
  pages = {100430},
  issn = {2451-9588},
  doi = {10.1016/j.chbr.2024.100430},
  urldate = {2025-08-14}
}

@misc{fangLeveragingAIGeneratedEmotional2025,
  title = {Leveraging {{AI-generated}} Emotional Self-Voice to Nudge People towards Their Ideal Selves},
  author = {Fang, Cathy Mengying and Chua, Phoebe and Chan, Samantha and Leong, Joanne and Bao, Andria and Maes, Pattie},
  year = 2025,
  month = apr,
  eprint = {2409.11531},
  primaryclass = {cs},
  doi = {10.1145/3706598.3713359},
  urldate = {2025-07-09},
  archiveprefix = {arXiv}
}

@article{fanMultiEmbodimentDigitalHumans2017,
  title = {Multi-Embodiment of Digital Humans in Virtual Reality for Assisting Human-Centered Ergonomics Design},
  author = {Fan, Kevin and Murai, Akihiko and Miyata, Natsuki and Sugiura, Yuta and Tada, Mitsunori},
  year = 2017,
  month = oct,
  journal = {Augmented Human Research},
  volume = {2},
  number = {1},
  pages = {7},
  issn = {2365-4325},
  doi = {10.1007/s41133-017-0010-6},
  urldate = {2025-06-18},
  langid = {english}
}

@inproceedings{hoppe_there_2022,
  title = {There Is No First- or Third-Person View in Virtual Reality: {{Understanding}} the Perspective Continuum},
  shorttitle = {There Is No First- or Third-Person View in Virtual Reality},
  booktitle = {Proceedings of the 2022 {{CHI}} Conference on Human Factors in Computing Systems},
  author = {Hoppe, Matthias and Baumann, Andrea and Tamunjoh, Patrick Chofor and Machulla, Tonja-Katrin and Wo{\'z}niak, Pawe{\l} W. and Schmidt, Albrecht and Welsch, Robin},
  year = 2022,
  month = apr,
  series = {{{CHI}} '22},
  pages = {1--13},
  publisher = {Association for Computing Machinery},
  address = {New York, NY, USA},
  doi = {10.1145/3491102.3517447},
  urldate = {2025-06-18},
  isbn = {978-1-4503-9157-3}
}

@inproceedings{jakeschCoWritingOpinionatedLanguage2023,
  title = {Co-Writing with Opinionated Language Models Affects Users' Views},
  booktitle = {Proceedings of the 2023 {{CHI}} Conference on Human Factors in Computing Systems},
  author = {Jakesch, Maurice and Bhat, Advait and Buschek, Daniel and Zalmanson, Lior and Naaman, Mor},
  year = 2023,
  month = apr,
  pages = {1--15},
  publisher = {ACM},
  address = {Hamburg Germany},
  doi = {10.1145/3544548.3581196},
  urldate = {2025-07-24},
  eventtitle = {{{CHI}} '23: {{CHI}} Conference on Human Factors in Computing Systems},
  isbn = {978-1-4503-9421-5},
  langid = {english}
}

@article{kourosDigitalMirrorsAI2024,
  title = {Digital Mirrors: {{AI}} Companions and the Self},
  shorttitle = {Digital Mirrors},
  author = {Kouros, Theodoros and Papa, Venetia},
  year = 2024,
  month = oct,
  journal = {Societies},
  volume = {14},
  number = {10},
  pages = {200},
  issn = {2075-4698},
  doi = {10.3390/soc14100200},
  urldate = {2025-07-24},
  langid = {english}
}

@inproceedings{liAIShellUnderstanding2023a,
  title = {{{AI}} in the Shell: {{Towards}} an Understanding of Integrated Embodiment},
  shorttitle = {{{AI}} in the Shell},
  booktitle = {Extended Abstracts of the 2023 {{CHI}} Conference on Human Factors in Computing Systems},
  author = {Li, Zhuying and Huang, Tianze and Patibanda, Rakesh and Mueller, Florian},
  year = 2023,
  month = apr,
  pages = {1--7},
  publisher = {ACM},
  address = {Hamburg Germany},
  doi = {10.1145/3544549.3585867},
  urldate = {2026-01-16},
  eventtitle = {{{CHI}} '23: {{CHI}} Conference on Human Factors in Computing Systems},
  isbn = {978-1-4503-9422-2},
  langid = {english}
}

@article{Liang2025widespreadllms,
  title = {The Widespread Adoption of Large Language Model-Assisted Writing across Society},
  author = {Liang, Weixin and Zhang, Yaohui and Codreanu, Mihai and Wang, Jiayu and Cao, Hancheng and Zou, James},
  year = 2025,
  journal = {Patterns},
  volume = {6},
  number = {12},
  pages = {101366},
  issn = {2666-3899},
  doi = {10.1016/j.patter.2025.101366}
}

@inproceedings{muellerNextStepsHumanComputer2020,
  title = {Next Steps for Human-Computer Integration},
  booktitle = {Proceedings of the 2020 {{CHI}} Conference on Human Factors in Computing Systems},
  author = {Mueller, Florian Floyd and Lopes, Pedro and Strohmeier, Paul and Ju, Wendy and Seim, Caitlyn and Weigel, Martin and Nanayakkara, Suranga and Obrist, Marianna and Li, Zhuying and Delfa, Joseph and Nishida, Jun and Gerber, Elizabeth M. and Svanaes, Dag and Grudin, Jonathan and Greuter, Stefan and Kunze, Kai and Erickson, Thomas and Greenspan, Steven and Inami, Masahiko and Marshall, Joe and Reiterer, Harald and Wolf, Katrin and Meyer, Jochen and Schiphorst, Thecla and Wang, Dakuo and Maes, Pattie},
  year = 2020,
  month = apr,
  pages = {1--15},
  publisher = {ACM},
  address = {Honolulu HI USA},
  doi = {10.1145/3313831.3376242},
  urldate = {2025-07-29},
  eventtitle = {{{CHI}} '20: {{CHI}} Conference on Human Factors in Computing Systems},
  isbn = {978-1-4503-6708-0},
  langid = {english}
}

@article{naeemShouldWeDiscourage2024,
  title = {Should We Discourage {{AI}} Extension? {{Epistemic}} Responsibility and {{AI}}},
  shorttitle = {Should We Discourage {{AI}} Extension?},
  author = {Naeem, Hadeel and Hauser, Julian},
  year = 2024,
  month = jul,
  journal = {Philosophy \& Technology},
  volume = {37},
  number = {3},
  pages = {91},
  issn = {2210-5441},
  doi = {10.1007/s13347-024-00774-4},
  urldate = {2025-07-21},
  langid = {english}
}

@article{ohata_i_2022,
  title = {I Hear My Voice; Therefore {{I}} Spoke: {{The}} Sense of Agency over Speech Is Enhanced by Hearing One's Own Voice},
  shorttitle = {I Hear My Voice; Therefore {{I}} Spoke},
  author = {Ohata, Ryu and Asai, Tomohisa and Imaizumi, Shu and Imamizu, Hiroshi},
  year = 2022,
  month = aug,
  journal = {Psychological Science},
  volume = {33},
  number = {8},
  pages = {1226--1239},
  issn = {0956-7976, 1467-9280},
  doi = {10.1177/09567976211068880},
  urldate = {2026-01-19},
  langid = {english}
}

@inproceedings{otonoTransformingEffectsVisual2023,
  title = {I'm Transforming! {{Effects}} of Visual Transitions to Change of Avatar on the Sense of Embodiment in {{AR}}},
  booktitle = {2023 {{IEEE}} Conference Virtual Reality and {{3D}} User Interfaces ({{VR}})},
  author = {Otono, Riku and Genay, Ad{\'e}la{\"i}de and {Perusqu{\'i}a-Hern{\'a}ndez}, Monica and Isoyama, Naoya and Uchiyama, Hideaki and Hachet, Martin and L{\'e}cuyer, Anatole and Kiyokawa, Kiyoshi},
  year = 2023,
  month = mar,
  pages = {83--93},
  doi = {10.1109/VR55154.2023.00024},
  urldate = {2025-06-18},
  eventtitle = {2023 {{IEEE}} Conference Virtual Reality and {{3D}} User Interfaces ({{VR}})}
}

@inproceedings{miuraMultiSomaDistributedEmbodiment2021,
  title = {{{MultiSoma}}: {{Distributed Embodiment}} with {{Synchronized Behavior}} and {{Perception}}},
  shorttitle = {{{MultiSoma}}},
  booktitle = {Augmented {{Humans Conference}} 2021},
  author = {Miura, Reiji and Kasahara, Shunichi and Kitazaki, Michiteru and Verhulst, Adrien and Inami, Masahiko and Sugimoto, Maki},
  year = 2021,
  month = feb,
  pages = {1--9},
  publisher = {ACM},
  address = {Rovaniemi Finland},
  doi = {10.1145/3458709.3458878},
  urldate = {2026-01-21},
  isbn = {978-1-4503-8428-5},
  langid = {english}
}

@misc{pataranutapornFutureYouConversation2024,
  title = {Future You: A Conversation with an {{AI-generated}} Future Self Reduces Anxiety, Negative Emotions, and Increases Future Self-Continuity},
  shorttitle = {Future You},
  author = {Pataranutaporn, Pat and Winson, Kavin and Yin, Peggy and Lapapirojn, Auttasak and Ouppaphan, Pichayoot and Lertsutthiwong, Monchai and Maes, Pattie and Hershfield, Hal},
  year = 2024,
  month = oct,
  number = {arXiv:2405.12514},
  eprint = {2405.12514},
  primaryclass = {cs},
  publisher = {arXiv},
  doi = {10.48550/arXiv.2405.12514},
  urldate = {2025-07-09},
  archiveprefix = {arXiv}
}

@article{rahillEffectsAvatarPlayersimilarity2021,
  title = {Effects of {{Avatar}} Player-Similarity and Player-Construction on Gaming Performance},
  author = {Rahill, Katherine M. and Sebrechts, Marc M.},
  year = 2021,
  month = aug,
  journal = {Computers in Human Behavior Reports},
  volume = {4},
  pages = {100131},
  publisher = {Elsevier BV},
  issn = {2451-9588},
  doi = {10.1016/j.chbr.2021.100131},
  urldate = {2025-07-11},
  langid = {english}
}

\end{document}